\documentclass{article}
\usepackage{spconf,amsmath,graphicx}
\usepackage{amsmath}
\usepackage{epstopdf}
\usepackage{amssymb}
\usepackage{systeme}
\usepackage{color}
\usepackage{pgfplots}
\usepackage{bbm}

\graphicspath{ C:\Users\alraam0a\Desktop\Imperfect_LASSO }

\graphicspath{ {C:\Users\alraam0a\Desktop\Imperfect_LASSO} }

\long\def\comment#1{}

\newfont{\bbb}{msbm10 scaled 700}

\newfont{\bb}{msbm10 scaled 1100}

\newcommand{\gv}{{\bf g}}
\newcommand{\hv}{{\bf h}}

\newcommand{\rv}{{\bf r}}
\newcommand{\sv}{{\bf s}}

\newcommand{\uv}{{\bf u}}
\newcommand{\wv}{{\bf w}}
\newcommand{\vv}{{\bf v}}
\newcommand{\xv}{{\bf x}}
\newcommand{\yv}{{\bf y}}
\newcommand{\zv}{{\bf z}}

\newcommand{\Am}{{\bf A}}

\newcommand{\Gm}{{\bf G}}
\newcommand{\Hm}{{\bf H}}

\newcommand{\Omegam}{\hbox{\boldmath$\Omega$}}

\newtheorem{theorem}{Theorem}

\usepackage{algorithm}% http://ctan.org/pkg/algorithm
\usepackage{algpseudocode}% http://ctan.org/pkg/algorithmicx
\title{Precise Performance Analysis of the LASSO under Matrix Uncertainties}
\name{Ayed M. Alrashdi$^{1,2}$, Ismail Ben Atitallah$^{1}$, Tareq Y. Al-Naffouri$^{1}$, and Mohamed-Slim Alouini$^{1}$}
\address{\small {$^{1}$ Computer, Electrical, Mathematical Sciences and Engineering (CEMSE) Division}\\ 
 \small {King Abdullah University of Science and Technology (KAUST), Thuwal, Saudi Arabia }\\
\small {$^{2}$Electrical Engineering Department, University of Hail, Hail, Saudi Arabia}}
\begin{document}

%\ninept
%
\maketitle
\begin{abstract}
In this paper, we consider the problem of recovering an unknown sparse signal $\xv_0 \in \mathbb{R}^n$ from noisy linear measurements $\yv = \Hm \xv_0+ \zv \in \mathbb{R}^m$. A popular approach is to solve the $\ell_1$-norm regularized least squares problem which is known as the LASSO. In many practical situations, the measurement matrix $\Hm$ is not perfectely known and we only have a noisy version of it. We assume that the entries of the measurement matrix $\Hm$ and of the noise vector $\zv$ are iid Gaussian with zero mean and variances $1/n$ and $\sigma_{\zv}^2$. In this work, an imperfect measurement matrix is considered under which we precisely characterize the limiting behavior of the mean squared error and the probability of support recovery of the LASSO. The analysis is performed when the problem dimensions grow simultaneously to infinity at fixed rates. Numerical simulations validate the theoretical predictions derived in this paper.
\end{abstract}
\vspace{-0.6cm}
\section{Introduction}
\label{sec:intro}
\vspace{-0.4cm}
The Least Absolute Shrinkage and Selection Operator (LASSO) \cite{tibshirani1996regression} is a powerfull method to recover a \textit{k-sparse} unknown signal $\xv_0$ $\in \mathbb{R}^{n}$ from noisy linear measurements:
$\yv = \Hm \xv_{0} + \zv \in \mathbb{R}^{m}$, 
where $ \Hm \in \mathbb{R}^{m \times n}$ is the measurement matrix, and $\zv \in \mathbb{R}^{m}$ is the noise vector. In this paper, we assume that $\Hm$ is not perfectly known, and we only have a noisy version of it that is denoted by $\Am$.
%In this paper, we assume that $\Hm$ is not known perfectly, and we only have a noisy version of it $\Am = \gamma \Hm + \epsilon \Omegam$ where $\gamma>0, \epsilon>0 \in \mathbb{R}$ and $\Omegam$ is an unknown error matrix.
Then, the LASSO solves the following convex optimization proplem: 
\vspace{-0.25cm}
\begin{equation}
\label{LASSO_1}   
\hat{\xv}=  \text{arg} \ \underset{{\xv}}{\operatorname{\min}} \ \frac{1}{2}||  \yv - \Am \xv ||^{2} + \lambda || \xv ||_1,
\end{equation}
%\vspace{-0.15cm}
where $|| \cdot ||$ and $|| \cdot ||_1$ denote the $\ell_2$-norm and the $\ell_1$-norm respectively, and $\lambda \geq 0$ is the regularization parameter that balances between the deviation of $\Am \hat{\xv}$ from the observations $\yv$ on one side, and the sparsity of the solution as promoted by the $\ell_1$-norm on the other side.
Problems of the form of (\ref{LASSO_1}) have many different diverse applications in science and engineering such as image processing \cite{ting2009sparse}, machine learning \cite{bishop2006pattern}, wireless communications \cite{gui2012improved}, etc.. 
%and sensor networks \cite{}. 
%Besides the importance of the LASSO algorithm and its diverse application, another important problem is to analyze its performance. A natural measure of the performance of (\ref{}) is the squared-error $MSE : = \frac{1}{n} || \hat{\xv} - \xv_0 ||^2$. Another important quality metric is the cosine similarity which arises in for example the task of document retrieval, where angular distance is a better way than MSE.
The LASSO has been studied from different prespectives over the years.
In recent years, the asymptotic exact characterization of the estimation performance gained a lot of interest. General performance metrics have been introduced such as the mean squared error and the probability support recovery.
% One important asymptotic property of it that has been of a lot of intreset in recent years is the exact characterization of the estimation performance such as the mean squared error or more general metrics such as the support recovery.
%Another performace metric is the support recovery. 
The first well-known bounds on the estimation performance of the lasso were order-wise in nature \cite{candes2006stable, candes2007dantzig, bickel2009simultaneous, negahban2009unified}. The Approximate Message Passing (AMP) framework has been used in \cite{donoho2009message, bayati2011dynamics, bayati2012lasso} to derive precise asymptotic analysis of the LASSO performance under the assumptions of iid Gaussian sensing matrix $\Am$. A recently developed framework, that is based on the Convex Gaussian Min-max Theorem (CGMT) \cite{thrampoulidis2016precise}, has been used in a series of works to precisely evaluate the estimation performance of non-smooth regularized convex estimators under noisy iid Gaussian measurements (including the LASSO) 
\cite{thrampoulidis2016precise,thrampoulidis2015regularized, stojnic2013framework,thrampoulidis2015precise, abbasi2016general}. 
%\cite{thrampoulidis2016precise} - \cite{abbasi2016general}.

However, these results assume that the measurement matrix $\Am$ is perfectly known. In many practical applications it is reasonable to expect uncertainty in the linear measurement matrix $\Am$ due to, e.g., imperfections in the signal acquisition hardware, model mismatch, estimation errors \cite{rosenbaum2010sparse}.
In this paper, we consider the additive uncertainty model: $\Am = \sqrt{1-\epsilon^2} \Hm + \epsilon \Omegam,$ where $\Hm$ is known and $\Omegam$ is an unknown error matrix and $\epsilon^2\in [0,1)$ is the variance of the error. Such model is commonly used in communication theory and known as imperfect Channel State Information (CSI) \cite{zenaidi2016performance}.

In this work, we derive precise asymptotic predictions of the \textit{mean squared error} and the \textit{support recovery} of the LASSO under the presence of uncertainties in the measurement matrix that has iid Gaussian entries (both $\Hm$ and $\Omegam$ have iid Gaussian entries). The Gaussianity assumption of the entries of $\Am$ is met in a wide range of applications such as MIMO application for Rayleigh fading model.
The analysis is based on the CGMT framework and is performed when the problem dimensions $m$, $n$ and $k$ all grow simultaneously to infinity at fixed rates. Although our analysis is asymptotic in nature, numerical simulations show that our theoretical predictions are valid even for a few dozens of the problem dimensions.
%In this work, we derive precise asymptotic predictions of the MSE and the cosine similarity of the LASSO under the presence of uncertainties in the design matrix that has iid Gaussian entries. In all previos work that based on the CGMT, the design matrix was assumed to be known perfectly but such knownledge is usually not avaliable in practice. 
\vspace{-0.5cm}
%\subsection{System Model and Assumptions}
\section{Problem Setup}
\vspace{-0.2cm}
\subsection{Performance Metrics}
\vspace{-0.3cm}
Finding a good estimate is an application dependent, since different applications require different desired properties of $\hat{\xv}$. This results in a need for a variety of different performance metrics. Here we discuss some of them.\\
\textbf{Mean squared error (MSE)}: A natural and heavily used measure of performance is the reconstruction \textit{mean squared error}, which measures the deviation of $\hat{\xv}$ from the true signal $\xv_0$. Formally, the MSE is defined as MSE $:= \frac{1}{n}|| \hat{\xv} - \xv_0 ||^2$.\\
\textbf{Support Recovery}: In the problem of sparse recovery, a natural measure of performance that is used in many applications (e.g. parameter selection in regression, sparse approximation, structure estimation in graphical models \cite{wainwright2009sharp}) is the support recovery, which is defined as identifying whether an entry of $\xv_0$ is on the support (i.e. non-zero), or it is off the support (i.e. zero). The decison is based on the LASSO solution $\hat{\xv}$: we say the $i^{th}$ entry of $\hat{\xv}$ is on the support if $| \hat{\xv}_{i}| \geq \xi$, where $\xi > 0$ is a user-defined hard threshold on the entries on $\hat{\xv}.$ In Theorem \ref{LASSO_on/off}, we precisely predict the \textit{per-entry} rate of successful on-support and off-support recovery. Formaly, let
\begin{subequations}\label{supp}
\begin{align}
\Phi_{\xi,\text{on}}(\hat{\xv}) = \frac{1}{k} \sum_{i \in S(\xv_0)} \mathbbm{1}_{\{| \hat{\xv}_{i}| \geq \xi \}}\\
\Phi_{\xi,\text{off}}(\hat{\xv}) = \frac{1}{n-k} \sum_{i \notin S(\xv_0)} \mathbbm{1}_{\{| \hat{\xv}_{i}| \leq \xi \}},
\end{align}
\end{subequations}
where $\mathbbm{1}_{\{\mathcal{B} \}}$ is the indicator function of a set $\mathcal{B}$, and $S(\xv_0)$ is the support of $\xv_0$, i.e. the set of the non-zero entries of $\xv_0$.
%\subsubsection{Working Assumptions}
%\vspace{-0.44cm}
\subsection{Working Assumptions}
\label{w_A}
%\vspace{-0.2cm}
The unkown signal $\xv_{0} \in \mathbb{R}^{n}$ is a $k$-sparse signal, i.e. only $k$ of its entries are sampled iid from a distribution $ p_{X_0}$ which has zero mean and unit variance ($\mathbb{E}[X_0^2] = 1$), and the remaining entires are zeros. 
%Also, we assume $\mathbb{E}_{X_0 \sim p_{X_0}}[X_0^2] < \infty$.
For the measurement matrix $\Am$, we consider the following additive uncertainty model:
%\begin{equation}
$\Am = \gamma \Hm + \epsilon \Omegam,$ 
%\end{equation}
%with the following \textit{assumptions}:
%\begin{itemize}
where $\Hm, \Omegam \in \mathbb{R}^{m \times n}$ both have entries iid $\mathcal{N}(0,1/n)$,  
%$\Omegam \in \mathbb{R}^{m \times n}$ is an unknown error matrix with entries i.i.d. Gaussian $\mathcal{N}(0,1/n)$,
 and $\epsilon^2 \in [0,1)$ is the variance of the error such that $\gamma^2 +\epsilon^2 = 1$.
The noise vector $\zv \in \mathbb{R}^{m}$ has entries iid $\mathcal{N}(0,\sigma_{\zv}^2)$.
%then $\xv_0$ is a $\kappa n$-sparse vector on average. 
The analysis is performed when the system dimensions ($m$, $n$ and $k$) grow simultaneously large at fixed ratios:
% $\delta:=\lim_{n \to \infty} \frac{m}{n}$.
$\frac{m}{n} \longrightarrow \delta \in (0,\infty)$, and $\frac{k}{n} \longrightarrow \kappa \in (0,1).$ Under these settings, the Signal to Noise Ratio (SNR) becomes SNR :=$\kappa/\sigma_{\zv}^2.$
%The analysis is performed when the system dimensions ($m$ and $n$) grow simultaneously large at a fixed ratio $\delta:=\lim_{n \to \infty} \frac{m}{n}$.
%We assume that the measurement matrix $\Am$ is not perfectly known due to some uncertainties. We consider the following additive uncertainty model:
%%\begin{equation}
%$\Am = \gamma \Hm + \epsilon \Omegam,$ 
%%\end{equation}
%with the following \textit{assumptions}:
%%\begin{itemize}
%$\Hm, \Omegam \in \mathbb{R}^{m \times n}$ both have entries iid $\mathcal{N}(0,1/n)$,  
%%$\Omegam \in \mathbb{R}^{m \times n}$ is an unknown error matrix with entries i.i.d. Gaussian $\mathcal{N}(0,1/n)$,
 %and $\epsilon^2 \in [0,1)$ is the variance of the error such that $\gamma^2 +\epsilon^2 = 1$,
%$\zv \in \mathbb{R}^{m}$ has entries iid $\mathcal{N}(0,\sigma_{\zv}^2)$, 
%and $\xv_{0} \in \mathbb{R}^{n}$ has entries $\xv_{0,i} \overset{iid}{\sim} p_{X_0} = (1-\kappa) \cdot \delta_0 + \kappa \cdot q_{X_0}$, where $\delta_0$ is the Dirac delta function, and $\kappa \in (0,1)$ is the sparsity level, and $ q_{X_0}$ is the pdf of the non-zero elements of $\xv_0$ which has variance $\sigma_{\xv}^2$. 
%%then $\xv_0$ is a $\kappa n$-sparse vector on average. 
%Also, we assume $\mathbb{E}_{X_0 \sim p_{X_0}}[X_0^2] < \infty$.
%\end{itemize}
\vspace{-0.4cm}
\subsection{Notation}
\vspace{-0.2cm}
Throughout this paper, we use boldface letters to represent vectors and matrices. We use the standard notation $\mathbb{P[\cdot]}$ and $\mathbb{E[\cdot]}$ to denote
probability and expectation. We write $X \sim p_X$ to denote that a random variable $X$ has a probability density/mass function $p_X$. In particular, $ H \sim \mathcal{N}(\mu, \sigma^2)$ implies that $H$ has Gaussian distribution of mean $\mu$ and variance $\sigma^2$. $\phi(x)$ and $Q(x)$ denote the pdf of a standard normal distribution and its associated Q-function respectively. For $a, \lambda \in \mathbb{R}$, such that $\lambda > 0$, we define the following functions:\\
The soft-thresholding operator: $\eta(a;\lambda)$ = $\text{arg min}_{x}$ $\frac{1}{2}(x-a)^2 + \lambda |x|$, which can be written:
\vspace{-0.3cm}
\begin{equation}\label{soft_TH}
\eta(a ; \lambda) =
\begin{cases} 
         a - \lambda & ,\text{if}  \ a > \lambda \\      
					
			0 & ,\text{if} \  |a| \leq \lambda  \\
			
       a + \lambda  & ,\text{if} \  a < -\lambda.
\end{cases}
\end{equation} 
and its optimal value $e(a;\lambda) =\text{min}_{x} \frac{1}{2}(x-a)^2 + \lambda |x|$
\vspace{-0.3cm}
\begin{equation}
e(a ; \lambda) =
\begin{cases} 
      \lambda a - \frac{1}{2} \lambda^2  & ,\text{if}  \ a > \lambda \\      
					
			\frac{1}{2} a^2 & ,\text{if} \  |a| \leq \lambda   \\
			
      -\lambda a - \frac{1}{2} \lambda^2  & ,\text{if} \   a < -\lambda .
\end{cases}
\end{equation}
%\vspace{-0.2cm}
%We define the \textit{cosine similarity} between the true signal $\xv_0$ and its estimate $\hat{\xv}$ as: $\cos(\hat{\xv}, \xv_0) = \frac{\hat{\xv}^T \xv_0}{|| \hat{\xv} || \cdot || \xv_0 ||}$.
%For a sequence of random variables $\{\mathcal{X}_n\}_{n \in N}$, $\mathcal{X}_n = \mathcal{O}_{p}(1)$ denotes boundedness in probability. 
Finaly, we write ``$\overset{P}{\longrightarrow} $" to designate convergence in probability.
\vspace{-0.3cm}
\section{Main Results}
\label{sec:Results}
\vspace{-0.4cm}
This section summarizes our main results on the precise analysis of the mean squared error and the probability of support recovery of the LASSO.
\vspace{-0.2cm}
\begin{theorem}[LASSO MSE]\label{LASSO_mse}
Fix $\lambda > 0$, and let $\hat{\xv}$ be a minimizer of the LASSO problem in (\ref{LASSO_1}), where $\Am, \zv$ and $\xv_0$ satisfy the working assumptions of Section \ref{w_A}. Then it holds in probability:
%\lim_{n\to\infty} \frac{1}{n} || \hat{\xv} - \xv_0 ||^2 = \alpha_*.
\vspace{-0.3cm}
\begin{align}
&\lim_{n\to\infty} \frac{1}{n} || \hat{\xv} - \xv_0 ||^2 = \delta \tau_{*}^2 - \sigma_{\zv}^2 \nonumber \\
&+2(\gamma -1)\mathbb{E}_{\underset{H \sim \mathcal{N}(0,1)}{X_0 \sim p_{X_0}} } \biggr[\eta \biggr(\gamma X_0 + \tau_* H ; \frac{2 \lambda \tau_*}{\beta_*} \biggl)  X_0  \biggl] ,
\end{align}
where 
$(\tau_*,\beta_*)$ is the unique solution to the following:
\vspace{-0.3cm}
\begin{align} \label{optimal_t_b}
&\underset{\tau > 0}{\operatorname{\min}} \ \underset{\beta > 0}{\operatorname{\max}} \ D(\tau, \beta ):= \frac{ \beta \tau  }{2} (\delta -1) + \frac{ \beta \sigma_{\zv}^2 }{2 \tau} - \frac{ \beta^2}{4} \nonumber \\
&+ \frac{ \beta \epsilon^2  \kappa}{2 \tau}  
  + \frac{\beta}{\tau} \cdot \mathbb{E}_{X_0,H} \biggr[e \biggr(\gamma X_0 + \tau H ; \frac{2 \lambda \tau}{\beta} \biggl) \biggl].
\end{align}
\end{theorem}
\vspace{-0.1cm}
$\tau_*$ and $\beta_*$ can be efficiently computed by writing the first order optimality conditions, i.e. $\nabla_{(\tau,\beta)}  D(\tau, \beta ).$
The proof of Theorem \ref{LASSO_mse} is based on the CGMT framework and is deferred to Section \ref{Proof}.\\
%\textbf{Remark 1} (High SNR): Theorm \ref{LASSO_mse}... \\
%Another important metric for assessing the performace of the LASSO is the support recovery, for which we present  a new precise asymptotic theoretical prediction in the following theorem.
The following Theorem precisely characterizes the support recovery metrics introduced in (\ref{supp}).
\vspace{-0.15cm}
\begin{theorem}[Probability of support recovery]\label{LASSO_on/off}
Under the same settings of Theorem \ref{LASSO_mse} and for any fixed $\xi>0$, it holds in probability that:
\vspace{-0.4cm}
\begin{equation*}
\lim_{n\to\infty} \Phi_{\xi,on}(\hat{\xv}) = \mathbb{P} [\bigl | \eta (\gamma X_0 + \tau_* H ; \frac{2 \lambda \tau_*}{\beta_*} ) \bigr |   \geq \xi ], 
\end{equation*}
\vspace{-0.3cm}
and 
%\vspace{-0.3cm}
\begin{equation*}
\lim_{n\to\infty} \Phi_{\xi,\text{off}}(\hat{\xv}) = \mathbb{P} [ \bigl| \eta (\tau_* H ; \frac{2 \lambda \tau_*}{\beta_*} ) \bigr|   \leq \xi ] = 1 -2 Q \biggl(\frac{\xi}{\tau_*} + \frac{2 \lambda}{\beta_*}\biggr).
\end{equation*}
\end{theorem}
\vspace{-0.2cm}
The proof of Theorem \ref{LASSO_on/off} is also based on the CGMT and largely follows the proof of Theorem \ref{LASSO_mse} and is omitted for space limitations.
%\textbf{Remark 3} (High SNR): Theorm \ref{LASSO_on/off}\\
%\textbf{Remark 4} (Optimal Tuning): 
\vspace{-0.5cm}
\section{Numerical Results}
\begin{figure}
  % This file was created by matlab2tikz.
%
%The latest updates can be retrieved from
%  http://www.mathworks.com/matlabcentral/fileexchange/22022-matlab2tikz-matlab2tikz
%where you can also make suggestions and rate matlab2tikz.
%
\begin{tikzpicture}

\begin{axis}[%
                    width=0.3\textwidth,
       height=0.2\textwidth,
at={(1.011in,0.642in)},
scale only axis,
xmin=0,
xmax=6,
xlabel style={font=\color{white!15!black}},
xlabel={$\lambda$},
ymin=0,
ymax=1,
ylabel style={font=\color{white!15!black}},
ylabel={MSE},
axis background/.style={fill=white},
xmajorgrids,
ymajorgrids,
legend style={legend cell align=left, align=left, draw=white!15!black}
]
\addplot [color=red]
  table[row sep=crcr]{%
0.001	0.915639577606094\\
0.101	0.305411140524806\\
0.201	0.184458563621176\\
0.301	0.132608193482286\\
0.401	0.104185793857657\\
0.501	0.0867182978929176\\
0.601	0.075332322347293\\
0.701	0.0677066209534927\\
0.801	0.0625844233581252\\
0.901	0.0592205346356877\\
1.001	0.0571415966634361\\
1.101	0.05602994662991\\
1.201	0.0556624302837552\\
1.301	0.0558759858452166\\
1.401	0.0565472518744471\\
1.501	0.0575799739193604\\
1.601	0.0588969666515019\\
1.701	0.0604348668607567\\
1.801	0.0621406333539185\\
1.901	0.063969201142702\\
2.001	0.065881892486469\\
2.101	0.0678453379749268\\
2.201	0.0698307373496554\\
2.301	0.0718133413673061\\
2.401	0.073772070966733\\
2.501	0.0756892148173288\\
2.601	0.0775501654195105\\
2.701	0.0793431688300926\\
2.801	0.081059074259318\\
2.901	0.0826910792586391\\
3.001	0.0842344733554255\\
3.101	0.0856863750847849\\
3.201	0.0870454855717882\\
3.301	0.0883118461786135\\
3.401	0.0894866156036861\\
3.501	0.0905718661330541\\
3.601	0.0915704012616666\\
3.701	0.0924855939340906\\
3.801	0.0933212479561066\\
3.901	0.0940814757529212\\
4.001	0.0947705960585943\\
4.101	0.0953930463504133\\
4.201	0.0959533091948541\\
4.301	0.0964558505644654\\
4.401	0.0969050685793484\\
4.501	0.0973052513335191\\
4.601	0.0976605426830549\\
4.701	0.097974915056433\\
4.801	0.0982521484915638\\
4.901	0.098495815213748\\
5.001	0.0987092691478087\\
5.101	0.0988956397881312\\
5.201	0.0990578299583209\\
5.301	0.0991985168676531\\
5.401	0.0993201560739775\\
5.501	0.0994249878348585\\
5.601	0.099515045419964\\
5.701	0.0995921649693497\\
5.801	0.099657996502447\\
5.901	0.0997140157276164\\
};
\addlegendentry{Theory}

\addplot [color=blue, draw=none, mark=o, mark options={solid,blue}]
  table[row sep=crcr]{%
0.001	0.955323164737541\\
%0.101	0.295333555033693\\
%0.201	0.179217488851335\\
0.301	0.128594617126165\\
%0.401	0.10787154852004\\
%0.501	0.0838166977690497\\
0.601	0.0775348815115045\\
%0.701	0.0741750586455607\\
%0.801	0.0617981337652087\\
0.901	0.0590807096362323\\
%1.001	0.0599618761483698\\
%1.101	0.0523877681494368\\
1.201	0.0519631227819481\\
%1.301	0.0569648403499048\\
%1.401	0.0534581724400643\\
1.501	0.0602303547675108\\
%1.601	0.0590218858779705\\
%1.701	0.0580783398435027\\
1.801	0.064194328677382\\
%1.901	0.0648200804652338\\
%2.001	0.0663427667025706\\
2.101	0.0664726094280378\\
%2.201	0.0684038651069493\\
%2.301	0.0744522828971224\\
2.401	0.0709208896928669\\
%2.501	0.0820880732624931\\
%2.601	0.0757464768891051\\
2.701	0.0769563965223018\\
%2.801	0.0741707463865082\\
%2.901	0.0846625010003924\\
3.001	0.0869106207730182\\
%3.101	0.0835355856997597\\
%3.201	0.0767766664562431\\
3.301	0.090188223695267\\
%3.401	0.0913379584601525\\
%3.501	0.0983119022631425\\
3.601	0.0949878794274334\\
%3.701	0.087327528136878\\
%3.801	0.0926633278611454\\
%3.901	0.0822545307552496\\
4.001	0.10067198945995\\
%4.101	0.111860216399786\\
%4.201	0.0970977402723565\\
4.301	0.104254042203568\\
%4.401	0.11261253857297\\
%4.501	0.0955236741216525\\
4.601	0.0961482076100996\\
%4.701	0.0875571687871322\\
%4.801	0.0928884235656775\\
4.901	0.0972855213680237\\
%5.001	0.0901228971228734\\
%5.101	0.100700792972886\\
5.201	0.0967214927088969\\
%5.301	0.10245884621681\\
%5.401	0.0937421487978371\\
5.501	0.0981106736497186\\
%5.601	0.0872331645817381\\
%5.701	0.100543038254335\\
5.801	0.0961908186418255\\
%5.901	0.0865633288729754\\
};
\addlegendentry{Simulation}

\end{axis}
\end{tikzpicture}%
\caption{\scriptsize {The MSE performance of the LASSO. Theoretical prediction from Theorem \ref{LASSO_mse}. For simulations $\kappa =0.1, \epsilon^2= 0.1,\delta = 0.8,n=256$, SNR = 0.5, and the data are averaged over 50 independent realizations of problem.}}%
\label{mse_Fig}
\end{figure}
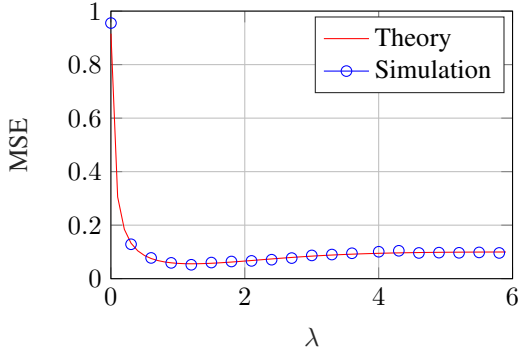

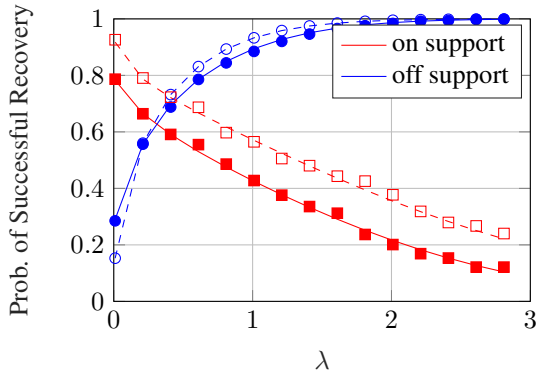
\begin{figure}
  \begin{tikzpicture}

\begin{axis}[%
  %               width=0.46\textwidth,
   %    height=0.4\textwidth,
                    width=0.4\textwidth,
       height=0.3\textwidth,
xmin=0,
xmax=3,
xlabel style={font=\color{white!15!black}},
xlabel={$\lambda$},
ymin=0,
ymax=1,
ylabel style={font=\color{white!15!black}},
ylabel={Prob. of Successful Recovery},
axis background/.style={fill=white},
xmajorgrids,
ymajorgrids,
legend style={legend cell align=left, align=left, draw=white!15!black}
]
\addplot [color=red]
  table[row sep=crcr]{%
0.01	0.782710090970864\\
%0.11	0.712130425518594\\
0.21	0.664381044755438\\
%0.31	0.625332902345487\\
0.41	0.590936582288211\\
%0.51	0.55947336854578\\
0.61	0.530051582179431\\
%0.71	0.502152943535689\\
0.81	0.475455200869464\\
%0.91	0.449750409141324\\
1.01	0.424902693991863\\
%1.11	0.400824433830763\\
1.21	0.377461829251375\\
%1.31	0.354785578765931\\
1.41	0.332784466423219\\
%1.51	0.311460666661119\\
1.61	0.290826088220514\\
%1.71	0.270899364995842\\
1.81	0.251703270919415\\
%1.91	0.233262441879405\\
2.01	0.215601356024046\\
%2.11	0.198742567018641\\
2.21	0.182705208972171\\
%2.31	0.16750380032574\\
2.41	0.153147369963754\\
%2.51	0.139638915590323\\
2.61	0.126975186035887\\
%2.71	0.115146759842045\\
2.81	0.104138375943908\\
%2.91	0.0939294611351472\\
};
\addlegendentry{on support}

\addplot [color=blue]
  table[row sep=crcr]{%
0.01	0.290296444328179\\
%0.11	0.451553689643173\\
0.21	0.556748256928438\\
%0.31	0.634627667556766\\
0.41	0.695774004804538\\
%0.51	0.745336057573205\\
0.61	0.786288475987665\\
%0.71	0.820542476157855\\
0.81	0.849418133859483\\
%0.91	0.873876669979063\\
1.01	0.894646996276548\\
%1.11	0.912300069119267\\
1.21	0.927295174038624\\
%1.31	0.940010060540857\\
1.41	0.950761261162301\\
%1.51	0.95981816609956\\
1.61	0.967412970687368\\
%1.71	0.973747808392798\\
1.81	0.978999918351005\\
%1.91	0.983325420072373\\
2.01	0.986862097766967\\
%2.11	0.98973148780168\\
2.21	0.992040489148816\\
%2.31	0.993882663019675\\
2.41	0.995339345440734\\
%2.51	0.996480660790987\\
2.61	0.997366493497713\\
%2.71	0.998047449022366\\
2.81	0.998565814477374\\
%2.91	0.998956514185291\\
};
\addlegendentry{off support}

\addplot [color=blue, draw=none, mark=*, mark options={solid, fill=blue, blue}, forget plot]
  table[row sep=crcr]{%
0.01	0.285176060313033\\
%0.11	0.446559109975658\\
0.21	0.556478544128833\\
%0.31	0.625382630381315\\
0.41	0.687886867381968\\
%0.51	0.743061977086858\\
0.61	0.78560580091237\\
%0.71	0.812395121570092\\
0.81	0.844228508586575\\
%0.91	0.867166765414227\\
1.01	0.884875645693699\\
%1.11	0.905799279560487\\
1.21	0.920312881090666\\
%1.31	0.934651366493099\\
1.41	0.946139463889221\\
%1.51	0.955146698472643\\
1.61	0.965819402273203\\
%1.71	0.969313281376226\\
1.81	0.97583039579492\\
%1.91	0.980516962080361\\
2.01	0.98389905404394\\
%2.11	0.98612553709202\\
2.21	0.991728743507524\\
%2.31	0.993824313639971\\
2.41	0.994162559192162\\
%2.51	0.995296553745416\\
2.61	0.997833027629659\\
%2.71	0.997300862776062\\
2.81	0.998593719549262\\
%2.91	0.998684989998096\\
};
\addplot [color=blue, dashed, mark=o, mark options={solid, blue}, forget plot]
  table[row sep=crcr]{%
0.01	0.152855590198673\\
%0.11	0.414506342129664\\
0.21	0.560043433340099\\
%0.31	0.659138197056223\\
0.41	0.731967963047845\\
%0.51	0.787627500300303\\
0.61	0.831136923448458\\
%0.71	0.865613558457912\\
0.81	0.893151046462775\\
%0.91	0.915237315371251\\
1.01	0.932976014965162\\
%1.11	0.947213257417824\\
1.21	0.958614447831329\\
%1.31	0.967712977634139\\
1.41	0.974942244733247\\
%1.51	0.980657441534614\\
1.61	0.985150920638146\\
%1.71	0.988663473729696\\
1.81	0.991392980229121\\
%1.91	0.993501331329227\\
2.01	0.995120180219151\\
%2.11	0.996355841581968\\
2.21	0.997293522768609\\
%2.31	0.998000988383231\\
2.41	0.998531719116894\\
%2.51	0.998927608974852\\
2.61	0.999221241158952\\
%2.71	0.999437784190304\\
2.81	0.999596551958898\\
%2.91	0.999712272174557\\
};
\addplot [color=red, draw=none, mark=square*, mark options={solid, fill=red, red}, forget plot]
  table[row sep=crcr]{%
0.01	0.786618259580987\\
%0.11	0.719069897962328\\
0.21	0.664119275225199\\
%0.31	0.613894116742344\\
0.41	0.591020241434859\\
%0.51	0.562179441495227\\
0.61	0.554896369221288\\
%0.71	0.500074789154888\\
0.81	0.485962425552939\\
%0.91	0.450075530199076\\
1.01	0.427782795332594\\
%1.11	0.416069693418224\\
1.21	0.3760914873906\\
%1.31	0.378976645536187\\
1.41	0.335641298598695\\
%1.51	0.316931427163923\\
1.61	0.312133008980143\\
%1.71	0.267440795801015\\
1.81	0.236817942959869\\
%1.91	0.217082483315371\\
2.01	0.201130610154996\\
%2.11	0.204875969592483\\
2.21	0.168881815347878\\
%2.31	0.165381465875603\\
2.41	0.153481670706548\\
%2.51	0.132612676528399\\
2.61	0.121396920126017\\
%2.71	0.125802586202216\\
2.81	0.121359752924587\\
%2.91	0.0941685456197631\\
};
\addplot [color=red, draw=none, mark=square, mark options={solid, red}, forget plot]
  table[row sep=crcr]{%
0.01	0.9262074707374\\
%0.11	0.846437017198145\\
0.21	0.790665693634123\\
%0.31	0.747870172264248\\
0.41	0.722845708192992\\
%0.51	0.68286791230198\\
0.61	0.687053464349574\\
%0.71	0.643127953875902\\
0.81	0.597302253004434\\
%0.91	0.595804768810399\\
1.01	0.564893032991066\\
%1.11	0.55331207468664\\
1.21	0.50565383095667\\
%1.31	0.503052394064396\\
1.41	0.480343912284625\\
%1.51	0.490390373445529\\
1.61	0.443135200578891\\
%1.71	0.433805447593444\\
1.81	0.425675386467194\\
%1.91	0.360181082792488\\
2.01	0.377307567118804\\
%2.11	0.317406779230757\\
2.21	0.318517277148542\\
%2.31	0.312203683382785\\
2.41	0.279322204166404\\
%2.51	0.27784023232928\\
2.61	0.267021198646183\\
%2.71	0.254503484746487\\
2.81	0.240177097201551\\
%2.91	0.198309017061182\\
};
\addplot [color=red, dashed, forget plot]
  table[row sep=crcr]{%
0.01	0.917564289928796\\
%0.11	0.832123376415774\\
0.21	0.786844915219929\\
%0.31	0.751745638640914\\
0.41	0.721282684582603\\
%0.51	0.693411512002182\\
0.61	0.667148295865189\\
%0.71	0.641949200607991\\
0.81	0.617490057339157\\
%0.91	0.593571367831159\\
1.01	0.570071495314563\\
%1.11	0.546921042960377\\
1.21	0.524087428211228\\
%1.31	0.50156467573835\\
1.41	0.479366017263394\\
%1.51	0.457518100204919\\
1.61	0.436056227116969\\
%1.71	0.415020374278704\\
1.81	0.394451900988702\\
%1.91	0.374390926583142\\
2.01	0.354874360096088\\
%2.11	0.335934546515728\\
2.21	0.317598464548615\\
%2.31	0.299887386962815\\
2.41	0.282816902223836\\
%2.51	0.266397196057304\\
2.61	0.250633501204255\\
%2.71	0.23552663910787\\
2.81	0.221073594880869\\
%2.91	0.207268083876121\\
};
\end{axis}
\end{tikzpicture}%
\caption{\scriptsize {Probability of scucessful on-support and off-support entries for two problem setup.  The theoretical prediction (Solid and dashed lines) comes from Theorem \ref{LASSO_on/off}. For the simulations
(Squares and Circles), we used $n = 256$, SNR= 0.5,  $\xi= 10^{-3}, \kappa = 0.1,\epsilon^2 =0.2,$
and the data are averaged over 50 independent realizations of problem. For solid lines and squares and circles, we used $\delta= 0.8$, while for dashed lines and empty squares and circles $\delta= 1.2$.}
}%
\label{on_off_Fig}
\end{figure}

\vspace{-0.3cm}
For illustration, we focus only on the case where $\xv_0$, has enties that are sampled from sparse Bernoulli distribution. i.e. most of the entries of $\xv_0$ are zeros and few are equal to 1.
The mean squared error of the LASSO is predicted by Theorem \ref{LASSO_mse}, and the particular term $\mathbb{E} [e (\gamma X_0 + \tau H ; \chi)]$, for $\tau>0$ in (\ref{optimal_t_b}) can be expressed as:
\vspace{-0.3cm}
\begin{equation*}
\kappa \int e(\gamma  + \tau h; \chi) \phi(h) d h + (1 - \kappa) \int e(\tau h; \chi) \phi(h) d h.
\end{equation*}
Figure \ref{mse_Fig} shows the accuracy of the mean squared error of the LASSO as predicted by Theorem \ref{LASSO_mse}. \\
\textbf{Remark} (Optimal Tuning): from Figure \ref{mse_Fig}, we can see that there is a value of regularizer $\lambda$ for which the MSE is minimized.\\
The prediction of theorem \ref{LASSO_on/off} for the support recovery compaired with the numerical simulations is shown in Figure \ref{on_off_Fig}. For the on-support recovery, the term $\mathbb{P} [\bigl | \eta (\gamma X_0 + \tau_* H ; \frac{2 \lambda \tau_*}{\beta_*} ) \bigr |   \geq \xi ] = Q(\frac{\xi +\gamma}{\tau_*}+ \frac{2 \lambda}{\beta_*}) + Q(\frac{\xi - \gamma}{\tau_*}+ \frac{2 \lambda}{\beta_*}) $ for the sparse Bernoulli case.
 Both figures show the high accuracy of our predictions.
 %\begin{figure}[h] 
%\includegraphics[width=6cm, height =5cm]{mse_Gauss003.eps}%
%\caption{The MSE performance of the LASSO. Theoretical prediction from Theorem \ref{LASSO_mse}. For simulations $\kappa =0.1, \epsilon^2= 0.1,\delta = 0.8,n=256$, SNR = 0.5.}%
%\label{mse_Fig}
%\end{figure}
 %\begin{figure}[h]%
%\includegraphics[width=\columnwidth]{Gauss_Similar.eps}%
%\caption{$\xv$ with i.i.d. sparse Gaussian entries, $\kappa =0.1, \epsilon= 0.45,\delta = 0.7,n=500$}%
%\end{figure}

\vspace{-0.6cm}
\section{Proof Outline}
\label{Proof}
\vspace{-0.3cm}
%\begin{proof}
In this section, we provide a proof outline of Theorem \ref{LASSO_mse}. For clarity, the steps of the proof are in divided into different subsections.
\vspace{-0.5cm}
\subsection{Convex Gaussian Min-max Theorem (CGMT)}
\vspace{-0.25cm}
We first need to state the key ingredient of the analysis which is the Convex Gaussian Min-max Theorem CGMT. Here, we just recall the statement of the theorem, and we refer the reader to \cite{thrampoulidis2016precise} for the complete technical requirements.
%The CGMT theorem associates with a primary optimization (PO) problem a simplified
%auxiliary optimization (AO) problem from which we can tightly infer properties of the original (PO), such
%as the optimal cost, the norm of the optimal solution. The (AO) problem is often easier to analyze because it does not involve big random matrices but only random vectors, in contrast to the (PO) which depends on the random design matrix $\Am$.
%Specifically, the (PO) and (AO) optimizations are given as follows:
Consider the following two min-max problems, which we refer to as the Primary Optimization (PO) and the Auxiliary Optimization (AO) problems:
\begin{subequations}
\begin{align}\label{P,AO}
&\Phi(\Gm) := \underset{\wv \in \mathcal{S}_{w}}{\operatorname{\min}}  \ \underset{\uv \in \mathcal{S}_{u}}{\operatorname{\max}} \ \uv^{T} \Gm \wv + \psi( \wv, \uv), \\
&\phi(\gv, \hv) := \underset{\wv \in \mathcal{S}_{w}}{\operatorname{\min}}  \ \underset{\uv \in \mathcal{S}_{u}}{\operatorname{\max}} \ || \wv || \gv^{T} \uv - || \uv || \hv^{T} \wv + \psi( \wv, \uv), \label{AA2}
\end{align}
\end{subequations}
where $\Gm \in \mathbb{R}^{m \times n}, \gv \in \mathbb{R}^{m}, \hv \in \mathbb{R}^n, \mathcal{S}_w \subset \mathbb{R}^n, \mathcal{S}_u \subset \mathbb{R}^m$ and $\psi : \mathbb{R}^n \times \mathbb{R}^m \mapsto \mathbb{R}$. Denote by $\wv_{\Phi} := \wv_{\Phi}(\Gm) $ and $\wv_{\phi} := \wv_{\phi}( \gv, \hv)$ any optimal minimizers of (\ref{P,AO}) and (\ref{AA2}) respectively. Let $\mathcal{S}_w, \mathcal{S}_u$ be convex, $\psi(\wv,\uv)$ be convex-concave continuous on $\mathcal{S}_w \times \mathcal{S}_u$, and $\Gm, \gv$ and $\hv $ all have \textit{iid} standard normal entries. Let $\mathcal{S}$ be any arbitrary open subset of $\mathcal{S}_w $.
Then, if $\lim_{n \rightarrow \infty} \mathbb{P}[\wv_{\phi} \in \mathcal{S}] = 1,$ it also holds $\lim_{ n \rightarrow \infty} \mathbb{P}[\wv_{\Phi} \in \mathcal{S}] = 1.$
%\begin{theorem}[CGMT \cite{DBLP:journals/corr/ThrampoulidisOH14a}]\label{TH:CGMT}
%\normalfont Let $\mathcal{S}_w, \mathcal{S}_u$ be convex, compact sets, and $\psi(\wv,\uv)$ be convex-concave continuous. For any open subset $\mathcal{S} \subset \mathcal{S}_w$:
%\begin{equation*}
%\lim_{n \rightarrow \infty} \mathbb{P}[\wv_{\phi} \in \mathcal{S}] = 1 \implies \lim_{ n \rightarrow \infty} \mathbb{P}[\wv_{\Phi} \in \mathcal{S}] = 1.
%\end{equation*}
%\end{theorem}
\vspace{-0.3cm}
\subsection{Identifying the (PO) and the (AO)}
\vspace{-0.3cm}
% Without loss of generality, we assume for the analysis that $\xv_0 = (1-\kappa) \delta_{0} + \kappa \hv$
For convenience, we consider the vector $\wv := \gamma \xv - \xv_0 $, then the problem in (\ref{LASSO_1}) can be reformulated in terms of $\wv$ as:
\vspace{-0.35cm}
\begin{equation}\label{Lasso_w}
\hat{\wv} = \text{arg} \ \underset{\wv}{\operatorname{\min}} \ || \Hm \wv + \frac{\epsilon}{\gamma} \Omegam(\wv+ \xv_0)-\zv ||^2 + \frac{2 \lambda}{\gamma} || \wv + \xv_0 ||_1.
\end{equation}
The problem in (\ref{Lasso_w}) is still not a form of a (PO) of the CGMT, so first we need to write it in form that suits the CGMT.
To do so, we first express the loss function of (\ref{Lasso_w}) in its dual form through the Fenchel conjugate, 
%\begin{equation*}
$|| \Hm \wv + \frac{\epsilon}{\gamma} \Omegam (\wv+ \xv_0)-\zv ||^2$
%\end{equation*}
%\begin{equation*}
$= \max_{\uv} \sqrt{n} \uv^T (\Hm \wv + \frac{\epsilon}{\gamma} \Omegam (\wv+ \xv_0)-\zv) - \frac{n}{4} || \uv ||^2$.
%\end{equation*}
The dual variable $\uv$ is scaled by a factor $\sqrt{n}$ to have a proper normalization that guarantees the convergence afterwards. 
%The dual variable $\uv$ is scaled by a factor $\sqrt{n}$  for an issue of convergence. 
%Strictly speaking, the terms in the objective function should be all of the same order $\mathcal{O}_p(n)$. 
Hence, the problem in (\ref{Lasso_w}) is equivalent to the following:
\vspace{-0.26cm}
\begin{align}\label{LASSO-PO1}
&\underset{\wv}{\operatorname{\min}} \ \underset{\uv}{\operatorname{\max}} \ \sqrt{n} \uv^T  \Hm \wv +\frac{\sqrt{n} \epsilon}{\gamma} \uv^T \Omegam (\wv+ \xv_0) -\sqrt{n} \uv^T \zv \nonumber \\
& - \frac{n}{4} || \uv ||^2 + \frac{2 \lambda}{\gamma} || \wv + \xv_0 ||_1.
\end{align}
The above problem is in the form of a (PO) of the CGMT. Therefore, we can define its corresponding (AO) as:
\vspace{-0.3cm}
\begin{align}\label{L_AO1}
&\underset{\wv}{\operatorname{\min}} \ \underset{\uv}{\operatorname{\max}} \ || \wv || \gv^T \uv - || \uv || \hv^T \wv +\frac{\sqrt{n} \epsilon}{\gamma} \uv^T \Omegam (\wv+ \xv_0) - \frac{n}{4} || \uv ||^2   \nonumber \\
& - \sqrt{n} \uv^T \zv + \frac{2 \lambda}{\gamma} || \wv + \xv_0 ||_1. 
\end{align}
\vspace{-0.95cm}
\subsection{Simplifying the (AO)} 
\vspace{-0.25cm}
The next step is to show that the (AO1) as it appears in (\ref{L_AO1}) can be transformed to a Scalar Optimization (SO) problem.
Since the vectors $\gv$ and $\hv$ are independent, 
$  || \wv || \gv^T \uv - \sqrt{n} \uv^T \zv   \buildrel d\over= \sqrt{|| \wv ||^2 + n \sigma_{\zv}^2}\gv^T \uv$. Therefore, (\ref{L_AO1}) is equivalent to
\begin{align}\label{simpleAO1}
&\underset{\wv}{\operatorname{\min}} \ \underset{\uv}{\operatorname{\max}} \ \sqrt{|| \wv ||^2 + n \sigma_{\zv}^2}\gv^T \uv - || \uv || \hv^T \wv - \frac{n}{4} || \uv ||^2  \nonumber \\
 &  +\frac{\sqrt{n} \epsilon}{\gamma} \uv^T \Omegam (\wv+ \xv_0) + \frac{2 \lambda}{\gamma} || \wv + \xv_0 ||_1.
\end{align}
Now, it is more convenient to work with $\xv$ instead of $\wv$,
\begin{align}\label{PO22}
&\underset{\xv}{\operatorname{\min}} \ \underset{\uv}{\operatorname{\max}}    \sqrt{n} \epsilon \uv^T \Omegam \xv + \sqrt{|| \gamma \xv - \xv_0 ||^2 + n \sigma_{\zv}^2}\gv^T \uv \nonumber \\
& - || \uv || \hv^T (\gamma \xv - \xv_0) - \frac{n}{4} || \uv ||^2 + 2 \lambda || \xv ||_1.
\end{align}
%\subsection{From (FAO) to (SAO)}
The optimization problem in (\ref{PO22}) can be seen as another primary optimization problem (PO2). Hence, we can define another auxiliary optimization problem (AO2) that corresponds to the new (PO2). First, let $\rv \in \mathbb{R}^m$ and $\sv \in \mathbb{R}^n$ be standard Gaussian vectors, then the (AO2) can be defined as:
\vspace{-0.25cm}
\begin{align}\label{AO2}
& \underset{\xv}{\operatorname{\min}} \ \underset{\uv}{\operatorname{\max}} \  \epsilon || \xv || \rv^{T} \uv - \epsilon  || \uv || \sv^{T} \xv  
+ \sqrt{|| \gamma \xv - \xv_0 ||^2 + n \sigma_{\zv}^2}\gv^T \uv \nonumber \\
& - || \uv || \hv^T (\gamma \xv - \xv_0) - \frac{n}{4} || \uv ||^2 + 2 \lambda || \xv ||_1.
\end{align}
Since $\rv$ and $\gv$ are independent standard Gaussian vectors, with abuse of notation, we have the following:
\vspace{-0.15cm}
\begin{equation*}
 \epsilon || \xv || \rv^{T} \uv + \sqrt{|| \gamma \xv - \xv_0 ||^2 + n \sigma_{\zv}^2}\gv^T \uv 
\end{equation*} 
%\begin{equation*}
%\buildrel d\over= \sqrt{|| \gamma \xv - \xv_0 ||^2 + n \sigma_{\zv}^2 + \epsilon^2 || \xv ||^2}\gv^T \uv
%\end{equation*}
\begin{equation*}
\buildrel d\over=  \sqrt{ || \xv ||^2 + || \xv_0 ||^2 -2\gamma \xv_{0}^{T} \xv + n \sigma_{\zv}^2   }\gv^T \uv.
\end{equation*}
Therefore, the (AO2) becomes:
\vspace{-0.2cm}
\begin{align}\label{firstAO2}
&\underset{\xv}{\operatorname{\min}} \ \underset{\uv}{\operatorname{\max}} \ \sqrt{ || \xv ||^2 + || \xv_0 ||^2 -2\gamma \xv_{0}^{T} \xv + n \sigma_{\zv}^2   } \gv^T \uv - \frac{n}{4} || \uv ||^2 \nonumber \\
& - || \uv|| (\epsilon \sv + \gamma \hv)^{T} \xv + ||\uv|| \hv^{T} \xv_0  + 2\lambda || \xv ||_1.
\end{align}
Fixing the norm of $\uv$ to $\beta: =|| \uv ||$, we can easily optimize over its direction by aligning it with $\gv$. Then the (AO2) simplifies to:
\vspace{-0.45cm}
\begin{align}
&\underset{\beta \geq 0}{\operatorname{\max}} \ \underset{\xv}{\operatorname{\min}} \ \sqrt{n} \beta \sqrt{ \frac{1}{n}(|| \xv ||^2 + || \xv_0 ||^2 -2\gamma \xv_{0}^{T} \xv) + \sigma_{\zv}^2} || \gv || \nonumber \\
& - \beta (\epsilon \sv + \gamma \hv)^{T} \xv + \beta \hv^{T} \xv_0  - \frac{n \beta^2}{4} + 2 \lambda || \xv ||_1.
\end{align}
To have a separable optimization problem, we use the following identity:
%\begin{equation}
$\sqrt{\chi} = \underset{\alpha > 0}{\operatorname{\min}} \ \frac{\alpha}{2} + \frac{\chi}{2 \alpha},$
%\end{equation}
where $\chi = \frac{1}{n}( ||\xv ||^2 + || \xv_0 ||^2 - 2 \gamma \xv_{0}^T \xv ) + \sigma_{\zv}^2$.
Also, define $\tau :=\frac{\sqrt{n} \alpha}{|| \gv ||}$, and $\tilde{\hv} := \epsilon \sv + \gamma \hv$. This yields the following optimization problem:
%\begin{align}\label{AA23}
%&\underset{\tau > 0}{\operatorname{\min}} \ \underset{\beta \geq 0}{\operatorname{\max}} \ \frac{ \beta \tau || \gv ||^2}{2} + \frac{n \beta \sigma_{\zv}^2 }{2 \tau} - \frac{n \beta^2}{4} + \frac{\beta || \xv_0 ||^2}{2 \tau}  \nonumber \\
% &+ \frac{\beta}{\gamma} (\tilde{\hv} - \epsilon \sv)^{T} \xv_0 + \sum_{i = 1}^{n} \biggl( \underset{\xv_i}{\operatorname{\min}} \ \frac{\beta}{2 \tau} \xv_i^2 \nonumber \\
%&- \beta \biggl( \tilde{\hv}_i + \frac{\gamma \xv_{0,i}}{\tau}  \biggr) \xv_i + 2 \lambda |\xv_i| \biggr). 
%\end{align}
%which can be further written as:
\begin{align}\label{AA23}
&\underset{\tau > 0}{\operatorname{\min}} \ \underset{\beta > 0}{\operatorname{\max}} \ \frac{ \beta \tau || \gv ||^2}{2} + \frac{n \beta \sigma_{\zv}^2 }{2 \tau} - \frac{n \beta^2}{4} 
 + \frac{\beta}{\gamma} (\tilde{\hv} - \epsilon \sv)^{T} \xv_0  \nonumber \\
 &+\frac{\beta}{\tau} \biggr( \sum_{i = 1}^{n} \frac{\epsilon^2}{2} \xv_{0,i}^2 - \gamma \tilde{\hv}_i \xv_{0,i} - \frac{\tau^2}{2} \tilde{\hv}_{i}^2 \biggl) \nonumber \\
&+ \frac{\beta}{\tau} \biggr( \sum_{i = 1}^{n} \underset{\xv_i}{\operatorname{\min}} \ \frac{1}{2} ( \xv_i - \gamma \xv_{0,i} - \tau \tilde{\hv}_i)^2 + \frac{2 \lambda \tau}{\beta} |\xv_i| \biggl).
\end{align}
The optimization over $\xv_i$ can be solved in a closed-form expression using the soft-thresholding operator, which is exactly the function defined in (\ref{soft_TH}).
%Also let $e(a;\lambda) =\text{min}_{x} \frac{1}{2}(x-a)^2 + \lambda |x|$, which is exactly the function defined in (\ref{}).
Then, the above optimization problem simplifies to the following Scalar Optimization (SO) problem:
 \vspace{-0.35cm}
%\begin{align}
%&\underset{\tau > 0}{\operatorname{\min}} \ \underset{\beta > 0}{\operatorname{\max}} \ \tilde{D}(\tau,\beta,\gv,\hv) := \frac{ \beta \tau || \gv ||^2}{2} + \frac{n \beta \sigma_{\zv}^2 }{2 \tau} - \frac{n \beta^2}{4} 
 % \nonumber \\
 %&+\frac{\beta}{\tau} \biggr( \sum_{i = 1}^{n} \frac{\epsilon^2}{2} \xv_{0,i}^2 - \gamma \hv_i \xv_{0,i} - \frac{\tau^2}{2} \hv_{i}^2 \biggl) \nonumber \\
%&+ \frac{\beta}{\gamma} \hv^{T} \xv_0 + \frac{\beta}{\tau} \sum_{i = 1}^{n} e \biggr( \gamma \xv_{0,i} + \tau \hv_i ; \frac{2 \lambda \tau}{\beta} \biggl).
%\end{align}
\begin{align}\label{AO22}
&\underset{\tau > 0}{\operatorname{\min}} \ \underset{\beta > 0}{\operatorname{\max}} \ \tilde{D}(\tau,\beta,\gv,\hv) := \frac{ \beta \tau || \gv ||^2}{2} + \frac{n \beta \sigma_{\zv}^2 }{2 \tau} - \frac{n \beta^2}{4} 
  \nonumber \\
 &+\frac{\beta}{\tau} \biggr( \sum_{i = 1}^{n} \frac{\epsilon^2}{2} \xv_{0,i}^2 - \gamma \hv_i \xv_{0,i} - \frac{\tau^2}{2} \hv_{i}^2 \biggl) \nonumber \\
&+ \frac{\beta}{\gamma} \hv^{T} \xv_0 + \frac{\beta}{\tau} \sum_{i = 1}^{n} e \biggr( \gamma \xv_{0,i} + \tau \hv_i ; \frac{2 \lambda \tau}{\beta} \biggl).
\end{align}
%in the above, recall $\tilde{\hv}$ definition and use $\hv$ instead of  $\tilde{\hv}$ .
\vspace{-0.95cm}
\subsection{Probabilistic asymptotic analysis of the (SO) problem}
\vspace{-0.25cm}
After simplifiying the (AO2) as in (\ref{AO22}), we are now in a position to analyze its limiting behavior.
First, we need to properly normalize the objective function in (\ref{AO22}) by dividing it by $n$. 
%This normalization is due to the fact that all the terms in (\ref{}) are of order $\mathcal{O}_p(n)$. 
Then, using the WLLN, we have: $\frac{1}{n} || \gv ||^2 \overset{P}{\longrightarrow} \delta, \frac{1}{n} || \hv ||^2 \overset{P}{\longrightarrow} 1$, $\frac{1}{n} || \xv_0||^2 \overset{P}{\longrightarrow} \kappa $ and $\frac{1}{n} \hv^T \xv_0 \overset{P}{\longrightarrow} 0$. Also, using the WLLN, it can be shown that for all $\tau>0$ and $\beta > 0$,
%\begin{equation*}
$\frac{1}{n} \sum_{i = 1}^{n} e ( \gamma \xv_{0,i} + \tau \hv_{i} ; \frac{2 \lambda \tau}{\beta} )\overset{P}{\longrightarrow} \mathbb{E} [ e ( \gamma X_{0} + \tau H ; \frac{2 \lambda \tau}{\beta} ) ]$, and $\frac{1}{n} \sum_{i=1}^{n} \tilde{\xv}_i \overset{P}{\longrightarrow} \mathbb{E}[ \eta (\gamma X_0 + \tau H ; \frac{2 \lambda \tau}{\beta})]$, where $\tilde{\xv}$ is the solution of (AO2) defined in (\ref{firstAO2}).
%\end{equation*}
Therefore, the point-wise convergence in $\tau$ and $\beta$ of the objective function in (\ref{AO22}) is the quantity $ D(\tau, \beta)$ defined in Theorem \ref{LASSO_mse}. Furthermore, it is possible to show that with probability one, the functions 
$\tau \mapsto {\operatorname{\max}}_{\beta>0} \tilde{D}(\tau,\beta,\gv, \hv)$ and $\tau \mapsto {\operatorname{\max}}_{\beta>0} D(\tau,\beta)$ are convex in $\tau$. Hence, it is possible to show using theorem 2.7 in \cite{newey1994large} that $\tau_n(\gv,\hv) \overset{P}{\longrightarrow} \tau_*$. 
%The latter convergence is cruial for the last step of the proof.
\vspace{-0.45cm}
\subsection{Applying the CGMT}
\vspace{-0.25cm} 
%For all $\tau>0$ and $\beta>0$, $\frac{1}{n} \sum_{i=1}^{n} \tilde{\xv}_i \overset{P}{\longrightarrow} \mathbb{E}[ \eta (\gamma X_0 + \tau H ; \frac{2 \lambda \tau}{\beta})]$, where $\tilde{\xv}$ is the solution of (AO2).
 We prove that the quantities $\hat{\xv} - \xv_0$ and $\tilde{\xv} - \xv_0$ are concentarted in the same set. Formally,
for any fixed $\zeta > 0$, we define the set:
$\mathcal{S} = \bigl \{ \vv: \bigl|  \frac{1}{n}|| \vv ||^2 - M(\tau_*,\beta_*)  \bigr| < \zeta \bigr\}$, where $M(\tau_*,\beta_*) = \delta \tau_*^2 - \sigma_{\zv}^2 +2(\gamma -1) \mathbb{E}[ \eta (\gamma X_0 + \tau_* H ; \frac{2 \lambda \tau_*}{\beta_*}) X_0]$, and $\tau_*$ and $\beta_*$ are as defined in Theorem \ref{LASSO_mse}.
%This cannot be done directly and we need to apply the CGMT twice. 
Let $\check{\xv}$ be the solution of (AO1) defined in (\ref{PO22}).
%Note that $||\tilde{\wv} || = || \gamma \tilde{\xv} - \xv_0 ||$, and the squared-error can be written as: $|| \tilde{\xv} - \xv_0 ||^2 = || \tilde{\wv} ||^2 + 2(\gamma - 1) \tilde{\xv}^T \xv_0$. Recall that $\tau_n(\gv,\hv) = \frac{\sqrt{n}}{||\gv||} \sqrt{\frac{1}{n} || \tilde{\wv} ||^2 + \sigma_{\zv}^2}$. Using $\tau_n(\gv,\hv)  \overset{P}{\longrightarrow} \tau_*$, we find $\frac{|| \tilde{\wv}||^2}{n} \overset{P}{\longrightarrow} \delta \tau_*^2 - \sigma_{\zv}^2$. 
The error can be written as: $|| \tilde{\xv} - \xv_0 ||^2 = || \tilde{\wv} ||^2 + 2(\gamma - 1) \tilde{\xv}^T \xv_0$. Recall that $\tau_n(\gv,\hv) = \frac{\sqrt{n}}{||\gv||} \sqrt{\frac{1}{n} || \tilde{\wv} ||^2 + \sigma_{\zv}^2}$. Using $\tau_n(\gv,\hv)  \overset{P}{\longrightarrow} \tau_*$, we find $\frac{|| \tilde{\wv}||^2}{n} \overset{P}{\longrightarrow} \delta \tau_*^2 - \sigma_{\zv}^2$. 
%For all $\tau>0, \beta> 0$, 
Also, it can be shown that $\frac{1}{n}\tilde{\xv}^T \xv_0 \overset{P}{\longrightarrow} \mathbb{E}[ \eta (\gamma X_0 + \tau_* H ; \frac{2 \lambda \tau_*}{\beta_*}) X_0]$. Putting all the results together, it can be shown that $\frac{1}{n} || \tilde{\xv} - \xv_0 ||^2 \overset{P}{\longrightarrow} M(\tau_*,\beta_*)$. This proves that for any $\zeta$, $\tilde{\xv} - \xv_0 \in \mathcal{S}$ with probability one. %w.p.1.
Then, we conclude using the CGMT that $\check{\xv} - \xv_0 \in \mathcal{S}$ with probability one. %w.p.1,
A second application of the CGMT is needed to conclude that $\hat{\xv} - \xv_0 \in \mathcal{S}$ with probability one and is omitted for space considerations.
This completes the proof of Theorem \ref{LASSO_mse}. 
\vspace{-0.4cm}
\section{Conclusion}
\vspace{-0.35cm}
In this paper, we proposed a precise asymptotic analysis of the MSE and the probability of support recovery of the LASSO under imperfect Gaussian measurement matrix assumptions. Although our analysis is asymptotic in nature, numerical simulations show that our theoretical predictions are valid even for a few dozens of the problem dimensions.
%Simulation results validate our theoretical predictions.
% even for small dimentions of the problem. 
%\section{Acknowledgment}
%The work of Ayed M. Alrashdi is supported by the University of Hail, Saudi Arabia.
\vfill\pagebreak

%\section{REFERENCES}
%\label{sec:refs}

% References should be produced using the bibtex program from suitable
% BiBTeX files (here: strings, refs, manuals). The IEEEbib.bst bibliography
% style file from IEEE produces unsorted bibliography list.
% -------------------------------------------------------------------------
\bibliographystyle{IEEEbib}
%\bibliography{strings,refs}
%\bibliography{strings}
\nocite{*} 
\bibliography{References}
\end{document}